# Nanosecond time-resolved dual-comb absorption spectroscopy

**Authors:** David A. Long[1*], Matthew J. Cich[2], Carl Mathurin[3], Adam T. Heiniger[2], Garrett C. Mathews[3], Augustine Frymire[3], Gregory B. Rieker[3]

**Affiliations:**

[1]National Institute of Standards and Technology, Gaithersburg, MD 20899, USA

[2]TOPTICA Photonics, Pittsford, NY 14534, USA

[3] Precision Laser Diagnostics Laboratory, University of Colorado, Boulder, CO 80309, USA

*Corresponding author D. A. Long. Email: david.long@nist.gov

**Abstract:**

Frequency combs have revolutionized the field of optical spectroscopy, enabling researchers to probe molecular systems with a multitude of accurate and precise optical frequencies. While there have been tremendous strides in direct frequency comb spectroscopy, these approaches have been unable to record high resolution spectra on the nanosecond timescale characteristic of many physiochemical processes. Here we demonstrate a new approach to optical frequency comb generation in which a pair of electro-optic combs is produced in the near-infrared and subsequently transferred with high mutual coherence and efficiency into the mid-infrared within a single optical parametric oscillator. The high power, mutual coherence, and agile repetition rates of these combs as well as the large mid-infrared absorption of many molecular species enable fully resolved spectral transitions to be recorded in timescales as short as 20 ns. We have applied this approach to study the rapid dynamics occurring within a supersonic pulsed jet, however we note that this method is widely applicable to fields such as chemical and quantum physics, atmospheric chemistry, combustion science, and biology.

**Main Text:**

**Time-resolved mid-infrared frequency comb spectroscopy has demonstrated great value for investigations of chemical and biological dynamics,[1-7] plasma discharges,[8,9] photolysis,[10-13] and supersonic/detonation systems.[14-17] The value derives from the large optical bandwidth, high resolution, and rapid spectral acquisition that is possible with optical frequency combs. Even still, the fastest high resolution mid-infrared absorption measurements have been limited to microsecond time resolution.[3,4,10,12,18-23] Achieving nanosecond time resolution, a characteristic timescale of many irreversible non-equilibrium and stochastic physicochemical processes, has proven challenging. Here we demonstrate an approach for nanosecond-timescale, high-resolution absorption spectroscopy in the mid-infrared region. A pair of near-infrared electro-optic-modulator frequency combs originating from the same single-frequency laser are simultaneously injected into a continuous-wave, singly resonant optical parametric oscillator. This dual comb pump generates a pair of mutually coherent mid-infrared frequency combs in a single collimated free-space beam. This configuration uniquely enables the generation of two mid-infrared frequency combs with arbitrary repetition rates, high power, high coherence, and relative simplicity, enabling quantitative measurements of absorption and spectral line-shapes at nanosecond timescales. Here we study the complex fluid dynamics of a supersonic jet expansion but note that this approach is amenable to studies of atmospheric re-entry, hypersonic engines, detonation processes, chemical kinetics, equilibrium dynamics, and other unsteady and non-repetitive processes.**

A schematic of the optical parametric oscillator (OPO)-based dual comb spectrometer can be found in Figure 1a. Light from a 1064 nm external-cavity diode laser is split onto two fibre arms, each passing through electro-optic phase modulators (EOMs) to generate two optical frequency combs with slightly different repetition rates. One beam also passes through an acousto-optic modulator to provide a relative frequency shift, $f_{AOM}$, ensuring that the resulting heterodyne beat frequencies for each pair of comb teeth are unique, $f_B = f_{AOM} + n\Delta f_{rep}$ (where $n$ is the integer comb tooth number and $\Delta f_{rep}$ is the difference in repetition rates). The pair of optical frequency combs are then combined and amplified from 5 mW to 10 W with an Yb fibre amplifier.

The amplified light is subsequently injected into a commercial continuous wave, singly resonant OPO (TOPTICA Photonics TOPO).[24,25] The OPO generates tunable mid-infrared radiation by down-converting near-infrared pump photons into near-infrared signal and mid-infrared idler photons with a periodically poled lithium niobate (PPLN) crystal. Because only the signal beam is resonant within the OPO, by energy conservation spectral structure on the pump, like the dual frequency combs, must be transferred exclusively to the non-resonant output, the mid-infrared idler.[26] In addition, the narrow linewidth of the pump laser is preserved on the signal and mid-infrared idler beams, where each of the beams had a linewidth below 50 kHz (see the Supplemental Material).

This approach possesses advantages compared to previous work in which mid-infrared optical frequency combs were generated by pumping a synchronous OPO with a near-infrared femtosecond laser.[27,28] In synchronous OPOs, a pulsed, mode locked pump laser generates a signal pulse which oscillates in the OPO cavity. The signal pulse must have a round trip time in the cavity that allows it to overlap with subsequent pump pulses in the nonlinear crystal, and this requires stabilization of the OPO cavity length. In the present method no cavity length stabilization is required because the signal beam is continuous wave, thus allowing for agile tuning of the infrared comb repetition rate. Further, this also allows for dual comb spectroscopy to be performed with a single OPO rather than requiring two tightly locked and separate OPOs.[29] A further comparison of the present approach to existing mid-infrared dual comb methods can be found in the Supplemental Material.

The present approach leverages the tunability and agility of EOM frequency combs, while producing mid-infrared optical powers >1 W which can be tuned from 2190 nm to 4000 nm. We note that these optical powers levels are three to five orders-of-magnitude higher than has been reached for mid-infrared electro-optic frequency combs based on other approaches such as difference frequency generation.[20,23,30] Figure 2 shows a resulting dual comb (multiheterodyne) interferogram (a) and a power spectrum (b). Without any active frequency or phase control, the heterodyne beat frequencies exhibit Fourier-transform-limited widths due to the common-mode nature of the dual comb generation from a single laser, even at a resolution as precise as 100 Hz. We note that this linewidth is far less than the kHz-order linewidth of the external cavity-diode laser. This result occurs because the two near-infrared combs were generated from the same laser leading to high mutual coherence – a property that was carried through into the mid-infrared. This configuration stands in contrast to dual comb approaches based on mode-locked frequency combs which require not only two separate frequency combs but also tight phase locking between them.[5,8,10,21,29]

Here we have applied this dual comb spectrometer to perform quantitative measurements of $CO_2$ within a supersonic jet. The jet system, shown in Figure 1b, consists of two 3.8 L pressure vessels as the plenum, a normally closed solenoid valve, a converging-diverging (CD) nozzle, and an optically accessible test section. A high-speed pressure transducer at the top of the test section monitors the gas pressure at the optical measurement location.

During testing, we initially evacuated the plenum to prevent dilution of the test gas and then filled the plenum with pure $CO_2$ to chamber pressures ranging from 138 kPa to 727 kPa. An automatic script opened the solenoid valve to produce a pulse of $CO_2$ which passed through the CD nozzle, into the test section, and then vented to the atmosphere. The CD nozzle produced choked flow at its throat and accelerated the fluid velocity to near Mach 1.8 at steady state when the plenum chamber pressure was greater than 185 kPa.

We leveraged the facile tuning of the OPO to measure $^{12}C^{16}O_2$ within the supersonic jet by probing its (10012) ← (00001) P26e transition at 3590.78 cm$^{-1}$. Typical spectra acquired with time resolutions of 20 ns and 1 µs can be found in Fig. 2c. We note that even at this resolution, the $CO_2$ transition can clearly be observed and more importantly its spectral area (and thus the sample mixing ratio) can be quantified.

Performing quantitative measurements at these fast rates is enabled by a unique combination of attributes of the present approach. Firstly, the agile repetition rates and modest bandwidths (near 30 GHz) of the electro-optic frequency combs allowed us to set $\Delta f_{rep}$ to values as large as 150 MHz while still ensuring all of the radiofrequency beatnotes occurred within the detector bandwidth. Thus, enabling us to extract individual comb teeth with Fourier transforms as short as 20 ns (see the Supplemental Material for further discussion). Secondly, the high power and coherence of these mid-infrared combs allowed for high signal-to-noise ratios even over very short timescales without signal averaging.

A typical time series of the integrated absorption of the P26e $CO_2$ transition during the supersonic jet pulse as well as the intracell pressure can be found in the upper panel of Fig. 3. We observed large oscillations of both the $CO_2$ absorption and the pressure which persisted for up to 40 ms after the pulse. Computational fluid dynamics (CFD) modelling of the pulsed jet system indicates that the oscillations in the measured integrated area are largely due to changes in the $CO_2$ mixing ratio resulting from oscillatory gas flow into and out of the test section once the valve is closed. As shown in Fig. 4, we also observed a clear dependence of this oscillatory behaviour on the chamber pressure which was also qualitatively observed in the CFD. We note that the CFD modelling was unable to capture all of the complexities of the experimental system (e.g., finite valve opening rates and a complete description of turbulent mixing), indicating the value of high speed, quantitative measurements especially in more computationally challenging applications such as detonation systems and hypersonic engines.

To further demonstrate the tunability and ultrahigh time resolution of the described method we have also applied it to record induced changes in absorption occurring at nanosecond timescales in a different gas. In this measurement the OPO output was split into two paths with one passing through a 20 cm long cell containing 1.33 kPa of $CH_4$ and 1.33 kPa of $C_2H_2$ and the second reference path bypassing the cell. The two beams were then combined on the photodetector. An acousto-optic modulator on the cell path was then employed to rapidly modulate the power in that path (measured 75% to 25% fall time of 170 ns) while the reference path power was held constant, thus leading to a change in the measured absorption. As shown in Fig. 5, the comb spectrometer is readily able to quantify this absorption change with a 100 ns time resolution. We note that these measurements of the 10011F2←00001A1 $CH_4$ transition at 4367.00 cm$^{-1}$ were performed nearly 500 nm from the $CO_2$ transition studied in the pulsed jet, with these large changes of the optical frequency comb wavelengths readily achieved with the OPO.

Unlike the realization of OPO-based frequency down conversion of femtosecond-based optical frequency combs, we found that our electro-optic combs do not require synchronous pumping of the OPO resonator. This important property allowed us to simultaneously produce a pair of high-power, mutually coherent mid-infrared frequency combs having different repetition rates in a single OPO. In this way the agility and flexibility of near-infrared electro-optical frequency combs can be efficiently transferred to the spectroscopically critical mid-infrared spectral region (where efficient, low $V_\pi$ electro-optic modulators are unavailable). For instance the use of modulators driven by tailored radiofrequency waveforms could provide ultra-dense frequency combs for ultrahigh resolution spectroscopy.[31] In addition, since calculations for this OPO yield a phase matching bandwidth as wide as several THz, the use of cascaded modulators and nonlinear spectral broadening techniques[32] could dramatically extend optical bandwidth. Further, there are potential quantum information applications if the OPO is operated below threshold to generate entangled frequency combs.[33]

We note that the recent availability of low-noise mid-infrared detectors with bandwidths of tens of gigahertz[34] should allow measurements of this type to reach single-nanosecond timescales and below while maintaining the present spectral resolution. In addition, the ability to record an isolated spectral line shape in a single shot should enable simultaneous measurements of quantities such as analyte mole fraction, temperature, density, and velocity in dynamic environments. The combination of frequency resolution, spectral power density, and acquisition speed will create new opportunities to measure process dynamics over a broad range of timescales with applications from protein folding to explosions.

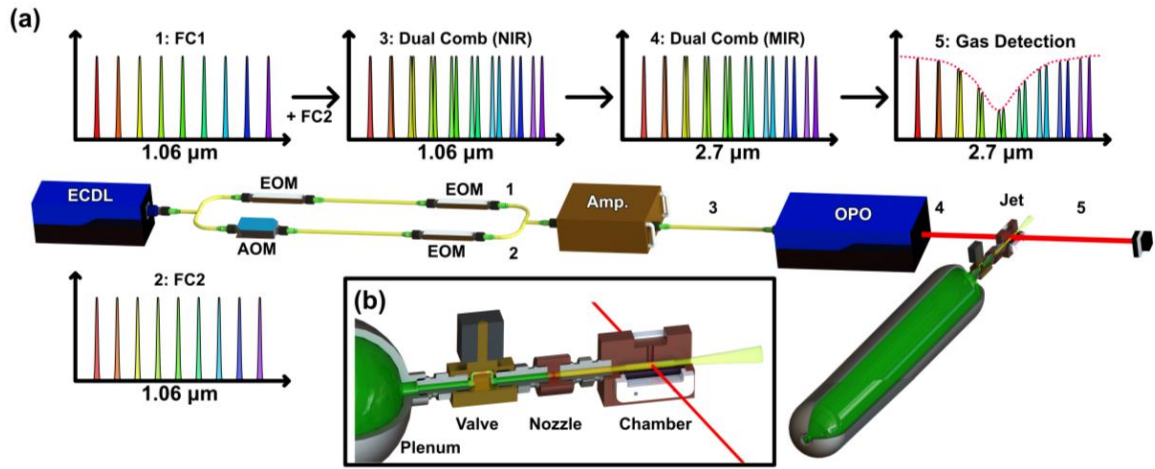

**Figure 1.** (a) The optical-parametric-oscillator-based, mid-infrared dual-comb spectrometer employed to interrogate a supersonic pulsed jet. A near-infrared external cavity diode laser (ECDL) is the common source for two optical frequency combs generated using electro-optic phase modulators (EOMs) having slightly different repetition rates. The two EOMs on the upper path are driven with identical waveforms to increase the modulation depth. The acousto-optic modulator (AOM) on the lower beam path shifts the beat frequencies between the two optical frequency combs away from DC. The pair of frequency combs are then combined and converted into mid-infrared wavelengths through the use of an amplifier (Amp.) and a singly resonant, continuous wave optical parametric oscillator (OPO). The idler beam is then passed through the supersonic jet and measured on a photodiode. All couplings shown in yellow are in fibre. (b) A machine drawing of the supersonic pulsed-jet showing the plenum, a normally closed solenoid valve, a converging-diverging (CD) nozzle, and the optical test chamber. The mid-infrared beam is shown in red.

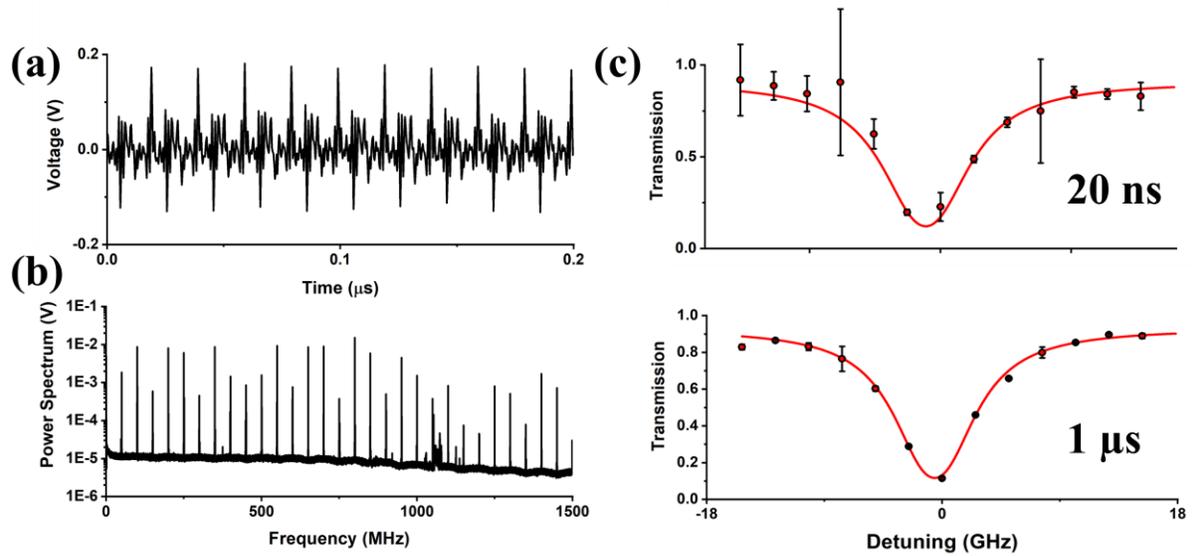

**Figure 2.** A portion of the observed mid-infrared optical frequency comb interferogram with a comb tooth spacing of 2.55 GHz and $\Delta f_{rep}$ = 150 MHz (a) and a power spectrum which was calculated as the average of 80 power spectra each comprising $1.5 \times 10^6$ samples (b). Panel (c) gives example $CO_2$ spectra recorded in 20 ns and 1 µs as well as the corresponding weighted Voigt profile fits with the frequency relative to the carrier frequency of 107 649 GHz. The uncertainties shown correspond to the relative standard deviation of the magnitudes of the individual measured comb teeth. The variation in the shown uncertainties is due to the varying comb tooth amplitudes (as can be seen in panel (b)). Importantly, the Fourier transform length (and thus the time resolution) can be selected after the data has been acquired.

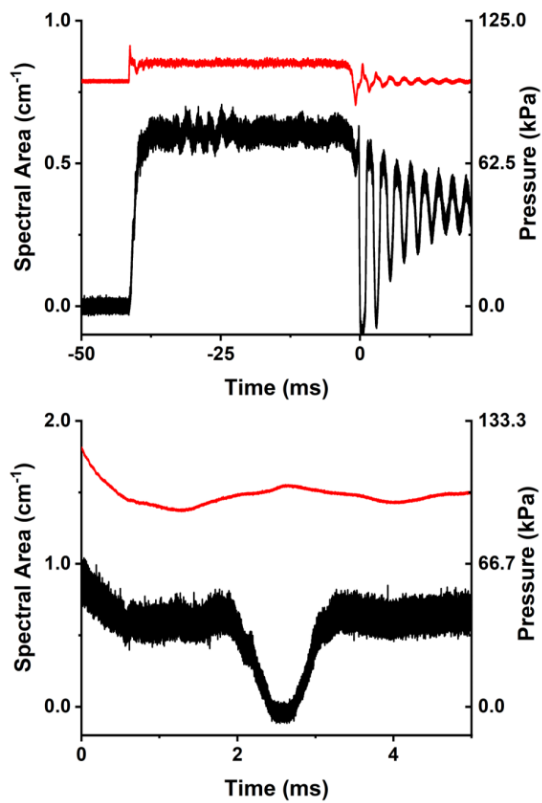

**Figure 3.** Measured change in the integrated absorption of the $CO_2$ P26e transition (black lines) during the pulsed jet and corresponding pressure trace (red lines). The upper panel shows the measurement at a chamber pressure of 138 kPa with a time resolution of 100 ns. The lower panel shows a portion of the traces after the pulse for a chamber pressure of 689 kPa with a time resolution of 20 ns. In both cases the mixing ratio and pressure measurements show differing behaviour, capturing complexity in the fluid mechanics occurring within the pulsed jet.

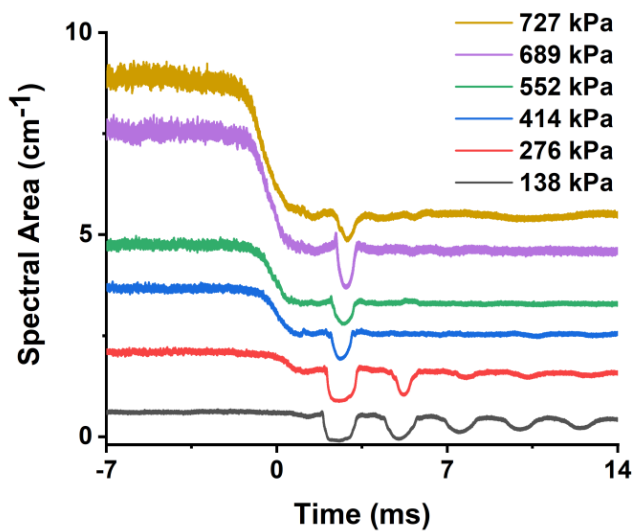

**Figure 4.** Observed $CO_2$ P26e transition integrated absorption during the falling edge of the pulsed jet for a range of chamber pressures. All traces have a time resolution of 100 ns and no averaging was performed. We observe a transition from the low-chamber pressure oscillatory regime to a single oscillation with increasing chamber pressures.

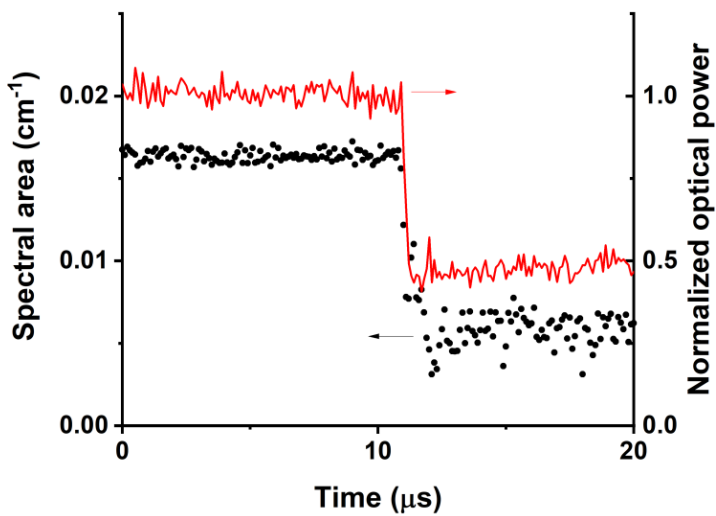

**Figure 5.** Observed $CH_4$ transition integrated absorption change (black points) induced by the modulation of the optical power on the cell leg (red line) while holding the reference path power constant. The integrated absorption was measured by the optical frequency comb spectrometer with a time resolution of 100 ns and no averaging was performed.

**References:**

1. Norahan, M. J. *et al.* Microsecond-resolved infrared spectroscopy on nonrepetitive protein reactions by applying caged compounds and quantum cascade laser frequency combs. *Anal. Chem.* **93**, 6779-6783 (2021).
2. Zhang, G., Horvath, R., Liu, D., Geiser, M. & Farooq, A. QCL-based dual-comb spectrometer for multi-species measurements at high temperatures and high pressures. *Sensors* **20**, 3602 (2020).
3. Pinkowski, N. H. *et al.* Dual-comb spectroscopy for high-temperature reaction kinetics. *Meas. Sci. Technol.* **31**, 055501 (2020).
4. Klocke, J. L. *et al.* Single-shot sub-microsecond mid-infrared spectroscopy on protein reactions with quantum cascade laser frequency combs. *Anal. Chem.* **90**, 10494-10500 (2018).
5. Draper, A. D. *et al.* Broadband dual-frequency comb spectroscopy in a rapid compression machine. *Opt. Express* **27**, 10814-10825 (2019).
6. Makowiecki, A. S. *et al.* Mid-infrared dual frequency comb spectroscopy for combustion analysis from 2.8 to 5 µm. *Proc. Combust. Inst.* **38**, 1627-1635 (2021).
7. Hoghooghi, N., Cole, R. K. & Rieker, G. B. 11-µs time-resolved, continuous dual-comb spectroscopy with spectrally filtered mode-locked frequency combs. *Appl. Phys. B* **127**, 17 (2021).
8. Abbas, M. A. *et al.* Broadband time-resolved absorption and dispersion spectroscopy of methane and ethane in a plasma Using a mid-infrared dual-comb spectrometer. *Sensors* **20**, 6831 (2020).
9. Krebbers, R. *et al.* Mid-infrared supercontinuum-based Fourier transform spectroscopy for plasma analysis. *Sci. Rep.* **12**, 9642 (2022).
10. Bjork, B. J. *et al.* Direct frequency comb measurement of OD + CO → DOCO kinetics. *Science* **354**, 444-448 (2016).
11. Yan, L. *et al.* Photolysis of Hi-CO nitrogenase – Observation of a plethora of distinct CO species using infrared spectroscopy. *Eur. J. Inorg. Chem.* **2011**, 2064-2074 (2011).
12. Agner, J. A. *et al.* High-resolution spectroscopic measurements of cold samples in supersonic beams using a QCL dual-comb spectrometer. *Mol. Phys.*, e2094297 (2022).
13. Fleisher, A. J. *et al.* Mid-Infrared Time-Resolved Frequency Comb Spectroscopy of Transient Free Radicals. *J. Phys. Chem. Lett.* **5**, 2241-2246 (2014).
14. Nair, A. P. *et al.* MHz laser absorption spectroscopy via diplexed RF modulation for pressure, temperature, and species in rotating detonation rocket flows. *Appl. Phys. B* **126**, 138 (2020).
15. Mathews, G. C., Blaisdell, M. G., Lemcherfi, A. I., Slabaugh, C. D. & Goldenstein, C. S. High-bandwidth absorption-spectroscopy measurements of temperature, pressure, CO, and $H_2O$ in the annulus of a rotating detonation rocket engine. *Appl. Phys. B* **127**, 165 (2021).
16. Nair, A. P., Minesi, N. Q., Jelloian, C., Kuenning, N. M. & Spearrin, R. M. Extended tuning of distributed-feedback lasers in a bias-tee circuit via waveform optimization for MHz-rate absorption spectroscopy. *Meas. Sci. Technol.* **33**, 105104 (2022).
17. Raza, M. *et al.* MHz-rate scanned-wavelength direct absorption spectroscopy using a distributed feedback diode laser at 2.3 µm. *Opt. Laser Technol.* **130**, 106344 (2020).
18. Pinkowski, N. H., Cassady, S. J., Strand, C. L. & Hanson, R. K. Quantum-cascade-laser-based dual-comb thermometry and speciation at high temperatures. *Meas. Sci. Technol.* **32**, 035501 (2020).

19  Sterczewski, L. A. *et al.* Cavity-enhanced Vernier spectroscopy with a chip-scale mid-infrared frequency comb. *ACS Photonics* **9**, 994-1001 (2022).
20  Jerez, B., Martín-Mateos, P., Walla, F., de Dios, C. & Acedo, P. Flexible electro-optic, single-crystal difference frequency generation architecture for ultrafast mid-infrared dual-comb spectroscopy. *ACS Photonics* **5**, 2348-2353 (2018).
21  Abbas, M. A. *et al.* Time-resolved mid-infrared dual-comb spectroscopy. *Sci. Rep.* **9**, 17247 (2019).
22  Luo, P.-L. & Chen, I. Y. Synchronized two-color time-resolved dual-comb spectroscopy for quantitative detection of HOx radicals formed from Criegee intermediates. *Anal. Chem.* **94**, 5752-5759 (2022).
23  Luo, P.-L., Horng, E.-C. & Guan, Y.-C. Fast molecular fingerprinting with a coherent, rapidly tunable dual-comb spectrometer near 3 μm. *Phys. Chem. Chem. Phys.* **21**, 18400-18405 (2019).
24  Foote, D. B. *et al.* High-resolution, broadly-tunable mid-IR spectroscopy using a continuous wave optical parametric oscillator. *Opt. Express* **29**, 5295-5303 (2021).
25  Henderson, A. & Stafford, R. Low threshold, singly-resonant CW OPO pumped by an all-fiber pump source. *Opt. Express* **14**, 767-772 (2006).
26  Chen, Y., Silfies, M. C., Kowzan, G., Bautista, J. M. & Allison, T. K. Tunable visible frequency combs from a Yb-fiber-laser-pumped optical parametric oscillator. *Appl. Phys. B* **125**, 81 (2019).
27  Leindecker, N., Marandi, A., Byer, R. L. & Vodopyanov, K. L. Broadband degenerate OPO for mid-infrared frequency comb generation. *Opt. Express* **19**, 6296-6302 (2011).
28  Bauer, C. P. *et al.* Dual-comb optical parametric oscillator in the mid-infrared based on a single free-running cavity. *Opt. Express* **30**, 19904-19921 (2022).
29  Muraviev, A. V., Smolski, V. O., Loparo, Z. E. & Vodopyanov, K. L. Massively parallel sensing of trace molecules and their isotopologues with broadband subharmonic mid-infrared frequency combs. *Nature Photon.* **12**, 209-214 (2018).
30  Yan, M. *et al.* Mid-infrared dual-comb spectroscopy with electro-optic modulators. *Light Sci. Appl.* **6**, e17076-e17076 (2017).
31  Long, D. A. & Reschovsky, B. J. Electro-optic frequency combs generated via direct digital synthesis applied to sub-Doppler spectroscopy. *OSA Continuum* **2**, 3576-3583 (2019).
32  Parriaux, A., Hammani, K. & Millot, G. Electro-optic frequency combs. *Adv. Opt. Photonics* **12**, 223-287 (2020).
33  Lu, Y. J. & Ou, Z. Y. Optical parametric oscillator far below threshold: Experiment versus theory. *Phys. Rev. A* **62**, 033804 (2000).
34  Dougakiuchi, T. & Akikusa, N. Application of High-Speed Quantum Cascade Detectors for Mid-Infrared, Broadband, High-Resolution Spectroscopy. *Sensors* **21**, 5706 (2021).
35  Levenberg, K. A method for the solution of certain non-linear problems in least squares. *Quart. Appl. Math.* **2**, 164 (1944).
36  Gordon, I. E. *et al.* The HITRAN2020 molecular spectroscopic database. *J. Quant. Spectrosc. Radiat. Transfer* **277**, 107949 (2022).

**Methods:**

*Optical parametric oscillator:* The OPO utilized in this work is a commercial OPO (TOPTICA TOPO) without any modifications made to the OPO aside from the introduction of modulation on the pump laser (as shown in Fig. 1a). A detailed description of the OPO design and tuning have been previously reported,[24,25] including a description of the OPO cavity geometry. The periodically-poled Lithium Niobate (PPLN) crystal is located inside an optical cavity which is resonant for signal photons only. The poling period of the PPLN varies along the crystal height in a fan-out structure. The cavity also contains an etalon which forces the signal to operate on only a single longitudinal mode with a single frequency narrowband spectrum. The cavity mirror reflectivities and PPLN length are similar to those found in Henderson et al.[25] We note that the measurement approach presented herein is not limited to this particular OPO, but rather would similarly apply to any singly-resonant, CW OPO.

The phase-matching bandwidth of the PPLN, the etalon free spectral range, and the etalon linewidth are chosen such that the cavity has high parametric gain only for a single longitudinal cavity mode. This results in a signal which is CW and single frequency. The signal spectrum is defined by the cavity and was measured to have < 50 kHz free-running linewidth during an 80 ms acquisition time (see the Supplemental Material for further linewidth measurements). The non-resonant idler has a spectrum which is the convolution of the pump and signal spectra. When the pump is a narrowband CW source, and the signal beam is at a single frequency, then by energy conservation the idler is also narrowband and CW. If the pump has spectral features that are broader than the signal linewidth, then these features will be replicated on the idler spectrum. In this work we avoid modulating the pump at rates which are synchronous with the cavity round trip time, which could result in multimode oscillation of the signal. However, even in cases where we had such modulation, we were still able to record high signal-to-noise spectra.

Translation of the crystal position relative to the pump beam changes the phase matching conditions, which widely tunes the signal (1450 nm to 2070 nm) and idler (2190 nm to 4000 nm) in discrete steps that correspond to the etalon free spectral range. Simultaneous translation of the crystal and rotation of the etalon improves coarse tuning resolution to < 30 GHz. We note that this tuning can be performed at rates up to 1 THz/s, however, tuning in excess of 300 GHz require a few seconds of thermal settling time. Continuous tuning of the idler center frequency is achieved by tuning the seed laser. Continuous tuning of the pump up to 300 GHz transfers to idler, and this can be performed in less than one second (pump tuning beyond 300 GHz causes a mode hop of the signal). This combination of tuning methods allows the idler rapidly to be brought within a few tens of MHz of a frequency of interest within seconds.

*Pulsed jet:* O-ring face seal fittings with internal diameters of 4.57 mm connected successive components of the pulsed-jet system to mitigate flow disruption at the mating surfaces. The normally closed solenoid valve blocked the flow at a 5.56 mm diameter internal orifice, and an electrical signal from a 120 V AC coil relay, controlled with a microcontroller board, opened the valve allowing gas to flow through the orifice. The converging-diverging nozzle constricted the pipe diameter from 4.57 mm to a throat diameter of 3.61 mm and expanded the diameter back to 4.57 mm with convergence and divergence half-angles of 10° and 6°, respectively. The test section comprises a 4.50 mm channel bracketed by two 2.5 mm thick sapphire windows. The idler beam passes through the test section at a 45.3° angle from normal (resulting in a measurement path length of 6.40 mm) on to a detector with 1.2 GHz bandwidth. The detector signal was then digitized at 3 gigasamples/s with a 12-bit digitizer. A 16-bit data acquisition system recorded the pressure transducer signals at 4.8 MS/s for the duration of the test.

*Data analysis:* Transmission spectra were calculated from measured interferogram signals and used to infer the spectral area of the $CO_2$ absorbance transition as a function of time. The raw interferogram was separated into sub-interferograms with durations corresponding to the desired time resolution (e.g., 20 ns or 100 ns). Discrete Fourier transforms of the individual interferograms were then performed to produce the radiofrequency-domain optical frequency comb spectra. Transmission spectra were then calculated by normalizing each measured comb spectrum by a comb spectrum acquired before the valve had opened. This normalization compensated for the intensity differences between individual comb teeth.

Spectral areas were quantified by fitting simulated transmission spectra to the measured spectra using a weighted nonlinear least squares fitting routine employing the Levenberg-Marquardt algorithm.[35] The uncertainty on each comb tooth, $\sigma$, was determined as the relative standard deviation of the comb tooth magnitudes during a representative set of spectra recorded with room air, and each datum contributing to the fit was weighted by $1/\sigma^2$. Transmission spectra were simulated using the Beer-Lambert law, given by:

$$\mathfrak{I}(\nu) = \exp[-A_{\text{int}}\, g(\nu; \nu_0, \Delta\nu_\text{D}, \Delta\nu_c) l\,]. \quad (1)$$

Here, $\mathfrak{I}(\nu)$ is the transmission at optical frequency $\nu$, $A_{\text{int}}$ is the integrated spectral area of the absorption transition, $g$ is the normalized Voigt line profile, $\nu_0$ is the transition frequency, $\Delta\nu_D$ is the Doppler full width, $\Delta\nu_c$ is the collisional (Lorentzian) full width, and $l$ is the optical pathlength. The transition frequency, Doppler width, and collisional width were held constant during the fits. The Doppler width was calculated as:

$$\Delta \nu_D = (7.1623 \times 10^{-7} \text{g mol}^{-1} \text{ K}^{-1})\, \nu_0 \sqrt{\frac{T}{M}}. \tag{2}$$

Here, $T$ is the temperature and $M$ is the molar mass of the absorbing species. The collisional-broadening width was calculated as:

$$\Delta \nu_c = 2\, P\, \gamma_{CO_2, T_0} \left(\frac{T_0}{T}\right)^N. \tag{3}$$

Here, $P$ is the pressure, $\gamma_{CO_2, T_0}$ is the self-broadening coefficient of $CO_2$ at the reference temperature $T_0 = 296$ K, and $N$ is the broadening coefficient temperature exponent. The broadening coefficient and temperature exponent were taken from the HITRAN2020 database.[36] In measurements with a plenum chamber pressure greater than 210 kPa, the pressure transducer saturated during the pulsed flow of $CO_2$, and an estimated pressure was used to calculate the collisional-broadening width. The pressure was estimated by assuming a linear relationship between the plenum chamber pressure and the average pressure measured during the quasi-steady duration of the pulsed flow. An estimated gas temperature of 196 K, calculated with isentropic flow relations assuming choked flow in the throat of the CD nozzle, was used to calculate the Doppler and collisional widths. The integrated spectral area was therefore the only free parameter during the fitting.

**Acknowledgments:** The authors thank Dr. Peter Hamlington, Tyler Souders, and Dr. Joseph T. Hodges for helpful discussions. The authors also acknowledge Dr. David Foote for his many contributions. Certain equipment, instruments, software, or materials are identified in this paper in order to specify the experimental procedure adequately. Such identification is not intended to imply recommendation or endorsement of any product or service by NIST, nor is it intended to imply that the materials or equipment identified are necessarily the best available for the purpose.

**Funding:** This material is based in part upon work supported by the Air Force Office of Scientific Research under award numbers FA9550-20-1-0328 and FA8649-20-0326.

**Author contributions:**

Conceptualization: DAL, MJC, ATH, GBR

Investigation: DAL, MJC, CM, GCM, AF, GBR

Funding acquisition: DAL, MJC, ATH, GBR

Writing – original draft: DAL, CM, ATH

Writing – review & editing: DAL, MJC, CM, GCM, ATH, GBR

**Competing interests:** The authors declare the following competing interests: MJC and ATH are employed by TOPTICA Photonics, Inc. TOPTICA Photonics, Inc. has submitted a patent related to this work and manufactured a portion of the equipment used in this work.

**Data and materials availability:** All data and supporting materials will be made available at a DOI to be provided through the National Institute of Standards and Technology.